-----------------------------------------------------------------------------

\documentstyle{elsart}
\begin{document}

\begin{frontmatter}
\rightline{CERN-TH/95-194}
\rightline{IFUM 511/FT}
\vskip 1truecm
\title{ Three-body  relativistic flux tube model \\
         from QCD Wilson-loop approach}

\author {N. Brambilla, G.M. Prosperi}
\address{Dipartimento di Fisica dell'Universit\`{a}, Milano, \\
INFN, Sezione di Milano, Via Celoria 16, 20133 Milano, Italy}
\author {and A. Vairo}
\address {Theory Division, CERN, CH-1211 Geneva 23, Switzerland, \\
INFN, Sezione di Bologna, Via Irnerio 46, 40126 Bologna, Italy}

\begin{abstract}
First we review the derivation of the relativistic flux tube model
for a quark-antiquark system from Wilson area law as we have given
in a preceding paper. Then we extend  the method to the three-quark
case and obtain a Lagrangian corresponding to a star flux tube
configuration.\par
A Hamiltonian can  be explicitly constructed as an expansion
in $1 / m^2$ or in the string tension $\sigma$. In the
first  case it reproduces the Wilson loop three-quark
semirelativistic  potential; in the second one, very complicated
in general,  but it reproduces  known string models for slowly
rotating quarks.
\end{abstract}

\end{frontmatter}

\vskip 4truecm
CERN-TH/95-194 \\
IFUM 511/FT \\
July 1995

\newpage

\pagenumbering{arabic}

\section{\bf Introduction}

In a recent paper  we have shown that, for the quark-antiquark
system, neglecting spin-dependent terms, it is possible
to derive rigorously  the so-called relativistic flux tube model
from the Wilson area law in QCD.\par
To achieve the result, the  essential ingredients were the path-integral
representation of the Pauli-type propagator we used in the derivation
of the semirelativistic potential, the replacement of the $1 / m^2$
expansion of the purely kinetic terms  by their closed  form and
the explicit integration over the momenta of the functional
integral.\par
In this letter we want to show  how a similar result
can be worked out for the three-quark system  starting from the
corresponding representation for the three-quark propagator we
have considered in ref. [1].
The resulting Lagrangian  turns out to correspond
to a star flux tube configuration. As in the quark-antiquark case,
the Hamiltonian can be  expressed only in the form of an expansion
in $1 / m^2$  or in the string tension $\sigma$. In the first
situation we obtain the Wilson loop semirelativistic potential,
but for the spin-dependent part. In the second one  we end up with a
very complicated result which, however, reduces to the three-quark
string model already considered in the literature for low angular
momentum. \par
In Section 2 we review the quark-antiquark case to clarify the procedure
and to establish notation; in Sec. 3 we derive the new Lagrangian and
in Sec. 4 we obtain the Hamiltonian of the resulting model.

\section{Quark-antiquark flux tube model}

In QCD the quark-antiquark  gauge invariant propagator can be written
\begin{eqnarray}
& & G(x_1,x_2,y_1,y_2) = \nonumber\\
& & \quad = \frac{1}{3}\langle0|{\rm T}\psi_2^c(x_2)U(x_2,x_1)\psi_1(x_1)
\overline{\psi}_1(y_1)U(y_1,y_2)  \overline{\psi}_2^c(y_2)
|0\rangle  =  \nonumber\\
& & \quad = \frac{1}{3} {\rm Tr} \langle U(x_2,x_1)
S_1^{{\rm F}}(x_1,y_1|A) U(y_1,y_2) C^{-1}
S_2^{{\rm F}}(y_2,x_2|A) C \rangle \> .
\end{eqnarray}
Here $c$ denotes the charge-conjugate fields, $C$ is the charge-conjugation
matrix, $U$ the path-ordered gauge string
\begin{equation}
U(b,a)= {\rm P}  \exp  \left(ig\int_a^b dx^{\mu} \, A_{\mu}(x)
\right) \>,
\end{equation}
(the integration path being the straight line joining $a$ to $b$),
 $S_1^{{\rm F}}$ and $S_2^{{\rm F}}$ the quark propagators in an
external gauge field $A^{\mu}$. Furthermore, in principle the angular
brackets should be defined as
\begin{equation}
\langle f[A] \rangle = \frac{\int {\cal D}A \, M_f(A)
f[A] e^{iS[A]}}
{\int {\cal D}A \, M_f(A) e^{iS[A]}} \> ,
\end{equation}
$S[A]$ being the pure gauge field action and $M_f(A)$
the determinant resulting from the explicit integration on the
fermionic fields. In practice in this paper  we
 take  $M_f(A)=1$ (quenched approximation).\par
{}From (1), by a Foldy-type transformation and with the other
elaborations described in ref. [1], after setting $y_1^0 = y_2^0=
t_{\rm i}$, $x_1^0=x_2^0 =t_{\rm f}$,
 we obtain the following Pauli-type
two-particle propagator
\begin{eqnarray}
& &K({\bf x}_1, {\bf x}_2;  {\bf y}_1, {\bf y}_2; t_{\rm f} - t_{\rm i}) =
\nonumber \\
& &\int_{{\bf y}_1}^{{\bf x}_1} {\cal D} {\bf z}_1 {\cal D}{\bf p}_2
\int_{{\bf y}_2}^{{\bf x}_2} {\cal D} {\bf z}_2 {\cal D} {\bf p}_2
\exp \Biggr\{i \Bigg[ \int_{t_{\rm i}}^{t_{\rm f}} dt \sum_{j=1}^2
\left({\bf p}_j
\cdot {\dot{\bf z}_j} - {{\bf p}^2_j \over 2 m_j} + {{\bf p}^4_j \over
8 m_j^3} +\dots \right) -\nonumber \\
 & &\quad \quad \quad \quad
-i  \ln W_{q \bar{q}} + {\rm spin} \, {\rm dependent } \, {\rm terms}
\,  \Bigg] \Biggr\} \> ,
\end{eqnarray}
$W_{q \bar{q}}$ being the so-called Wilson loop integral
\begin{equation}
W_{q\overline{q}} = \frac{1}{3} \left\langle {\rm Tr \, P}
 \exp \left( ig \oint_{\Gamma} dx^{\mu} \,
 A_{\mu}(x) \right) \right\rangle \> .
\end{equation}
Here the integration loop $\Gamma$ is assumed to be made by the quark
world line $\Gamma_1$, the antiquark world line $\Gamma_2$ (described
in the reverse direction) and the two straight lines which  connect
${\bf x}_1$ to ${\bf x}_2$, ${\bf y}_2$ to ${\bf y}_1$ and close
the contour (see Fig. 1). As usual
$A_{\mu} (x) = \frac{1}{2} {\lambda}_ {a} A_{\mu} ^{a} (x)$,
P prescribes the ordering of the colour matrices (from right to left)
according to the direction fixed on the loop, and the
``functional measures'' are  given by
\begin{equation}
{\cal D} {\bf z}   =  \left({m\over 2 \pi i \varepsilon }\right)^{3N\over 2}
d {\bf z}_1 \dots  d {\bf z}_{N-1} \> , \quad
{\cal D} {\bf p}   =  \left({i \varepsilon\over 2 \pi m}\right)^{3N\over 2}
d {\bf p}_1 \dots  d {\bf p}_{N-1} d {\bf p}_N \> ,
\label{eq:monodim}
\end{equation}
where $\varepsilon=t / N$, the limit $N\to \infty$ is understood
and the end points $x$ and $y$ stand for the conditions $z_0=y$, $z_N=x$. \par
After neglecting  the spin-dependent terms and reintegrating the
kinematical terms in their closed form, (4) becomes
\begin{eqnarray}
& &K({\bf x}_1, {\bf x}_2;  {\bf y}_1, {\bf y}_2; t_{\rm f} - t_{\rm i}) =
\int_{{\bf y}_1}^{{\bf x}_1} {\cal D} {\bf z}_1 {\cal D}{\bf p}_2
\int_{{\bf y}_2}^{{\bf x}_2} {\cal D} {\bf z}_2 {\cal D} {\bf p}_2
 \times \nonumber \\
& & \quad \quad \quad
\exp \left\{i \left[ \int_{t_{\rm i}}^{t_{\rm f}} dt \sum_{j=1}^2
\left({\bf p}_j \cdot {\dot{\bf z}_j} - \sqrt{m_j^2+ {\bf p}_j^2}
\right )  -i  \ln W_{q \bar{q}}
 \right ] \right\} \> .
\end{eqnarray}
As usual we assume that  $i \ln W_{q \bar{q}}$ can be written as the
sum of a perturbative term (at the lowest order) and an area
and a perimeter term
\begin{equation}
i  \ln W^{{\rm SR}}_{q\overline{q}} =
 \frac{4}{3} g^2 \int_{\Gamma_1}dx^{\mu}_1
 \int_{\Gamma_2} dx^{\nu}_2 \, iD_{\mu \nu} (x_1-x_2)
 + {\sigma} S_{\min} + C { P}  \> ,
\end{equation}
$S_{\rm min}$ being the minimal area enclosed by $\Gamma$,
$P$ its perimeter and $D_{\mu \nu}$ the usual gluon
propagator. Furthermore, to evaluate $S_{\rm min}$, we adopt
the {\it straight line approximation}, consisting of replacing
the minimal area  with the area spanned by equal time straight lines
joining the quark and the antiquark
\begin{equation}
S_{\rm min} = \int_{t_{\rm i}}^{t_{\rm f}} dt \int_0^1 ds \vert {\bf z}_1
-{\bf z}_2 \vert \left[1 - \left( s \dot{\bf z}_{1 {\rm T}}+ (1-s)
\dot{\bf z}_{2 {\rm T}} \right)^2 \right]^{1\over 2} \> ,
\end{equation}
having set $ \dot{z}_{j {\rm T}}^h = (\delta^{hk} - \hat{r}^h
\hat{r}^k ) \dot{z}_j^k$ (transversal velocity), $\hat{r}^h = r^h /r$ and
${\bf r}= {\bf z}_1- {\bf z}_2$.
We also use the Coulomb gauge and adopt the {\it instantaneous approximation}
on the perturbative term. So we can write
\begin{eqnarray}
& &i   \ln W_{ q \bar{q}} = -{4 \over 3}
{{\alpha}_s  \over r} \left[ 1 - \frac{1}{2}
( \delta^{hk} + \hat{r}^h \hat{r}^k ) \dot{z}_1^h \dot{z}_2^k \right] + \\
& & \quad  + \sigma
r \int_0^1 ds \left[1 - ( s \dot{\bf z}_{1{\rm T}} + (1-s) \dot{{\bf
z}}_{2{\rm T}} )^2 \right]^{1/2}
+ C \left( \sqrt{1- \dot{\bf z}_1^2} +\sqrt{1-\dot{\bf z}_2^2}
\right) \> .
\nonumber
\end{eqnarray}
Notice that in the first term in (8) and in the last in (10),
we have neglected the contributions coming
 from the end straight lines ($t=t_{\rm f}$ and $ t=t_{\rm i}$) having
 in mind the limit $t_{\rm f}-t_{\rm i} \to \infty$.
Furthermore, (9) not being  Lorentz invariant, our  assumptions are
understood to be made in the centre-of-mass frame.\par
Then we replace (10) in (7), expand the kinetic terms around ${\bf p}_j
= m_j \dot{\bf z}_j/$ $\sqrt{1- \dot{\bf z}_j^2}$
\begin{eqnarray}
{\bf p} \cdot \dot{z} - \sqrt{m^2 + {\bf p}^2 } & =&
 - m \sqrt{1 -\dot{\bf z}^2} - {1\over 2 m}
 ( 1- \dot{\bf z}^2)^{1\over 2} ( \delta^{hk} - \dot{z}^h \dot{z}^k)
\nonumber \\
 & & \left(p^h - { m\dot{z}^h \over \sqrt{1 -\dot{\bf z}^2}}
\right) \left( p^k - {
m \dot{z}^k \over \sqrt{1 -\dot{\bf z}^2}} \right) + \dots \> ,
\end{eqnarray}
and perform explicitly  the integration  over the momenta
neglecting the successive terms in (11) (Gaussian approximation).
We obtain
\begin{eqnarray}
& & K( {\bf x}_1 , {\bf x}_2; {\bf y}_1 ,{\bf y}_2;
 t_{\rm f} - t_{\rm i} ) =\nonumber \\
& &\quad \int_{{\bf y}_1}^{{\bf x}_1} {\cal D}  {\bf z}_1
 \Delta [ {\bf z}_1 ] \int_{{\bf y}_2}^{{\bf x}_2}
 {\cal D}  {\bf z}_2
\Delta [ {\bf z}_2 ] \exp \left\{i \int_{t_{\rm i} }^{t_{\rm f}}
dt L( {\bf z}_1 ,
{\bf z}_2 , \dot {\bf z}_1 , \dot {\bf z}_2 ) \right\} \> ,
\end{eqnarray}
with
\begin{eqnarray}
L \, & & = - \sum_{j=1}^2 m_j \sqrt{1- \dot {\bf z}_j^2 } +
 {4 \over 3} { {\alpha}_s  \over r} \left[ 1 - \frac{1}{2}
( \delta^{hk} + \hat{r}^h \hat{r}^k ) \dot{z}_1^h \dot{z}_2^k \right]
+\nonumber \\
 & & {} - \sigma
r \int_0^1 ds \left[1 - ( s \dot{\bf z}_{1{\rm T}} + (1-s) \dot{{\bf
z}}_{2{\rm T}} )^2 \right]^{1/2}- C \sum_{j=1}^2 \int_{t_{\rm i}}^{t_{\rm f}}
dt \sqrt{1-\dot{\bf z}_j^2 } \> ,
\end{eqnarray}
and
\begin{equation}
\Delta [ {\bf z} ]= \left\{ \prod_t \det \left[{1 \over m}
(1 - \dot{\bf z}^2)^{1/2} ( \delta^{hk}-{\dot{z}}^h {\dot{z}}^k ) \right]
\right\}^{-\frac{1}{2}} \> = \prod_t ( 1 -\dot{\bf z}^2 ) ^{-{3\over 4}} \> .
\end{equation}
Of course the factor  $\Delta [z]$ has to be considered part of the
relativistic ``functional measure'' in the configuration space.\par
Notice that if we introduce explicitly  the factor $1 / \hbar$
in front of the exponential  in (7), the result  expressed by
(12) and (13) becomes exact in the limit  $\hbar \to 0$.
This implies that (13) provides already  the exact classical
Lagrangian of the system, while to go beyond the  Gaussian approximation
would amount to modifying the expression of $\Delta({\bf z})$ alone.
As a matter of fact, the additional terms would be highly singular
in the time lattice spacing $\varepsilon$ and would  not match
in reconstructing a time integral  in the exponent (the fact can be
checked explicitly  performing the exact  integrals in the
momenta [2]). \par
The Lagrangian defined by (13) is the Lagrangian of
the relativistic flux tube model [3,4]  with the addition of a Coulombic
term. The perimeter term can be absorbed in a redefinition of the quark
masses ($m_j \to m_j + C$) and from here on we ignore it.\par
{}From (13) we can introduce the canonical momenta
\begin{eqnarray}
{\bf p}_1&= & {\partial L \over \partial \dot{\bf v}_1}=
{m_1 {\bf v}_1 \over \sqrt{1-{\bf v}_1^2}} + \sigma r
 \int_0^1 ds \, {s \,{\bf v}_{\rm t} \over
 \sqrt{1-{\bf v}_{\rm t}^2 } }  \> , \nonumber \\
{\bf p}_2 &=& {\partial L\over \partial \dot{\bf v}_2}
={m_2 {\bf v}_2 \over \sqrt{1-{\bf v}_2^2}} +
\sigma r
 \int_0^1 ds \, { (1-s)  {\bf v}_{\rm t} \over
 \sqrt{1-{\bf v}_{\rm t}^2} } \> ,
\end{eqnarray}
while the total linear momentum and the Hamiltonian turn out respectively
to be
\begin{equation}
{\bf P}= {\bf p}_1 +{\bf p}_2={m_1 {\bf v}_1 \over \sqrt{1-{\bf v}_1^2}}
+{m_2 {\bf v}_2 \over \sqrt{1-{\bf v}_2^2}}
+ \sigma r \int_0^1 ds  { {\bf v}_{\rm t} \over
 \sqrt{1-{\bf v}_{\rm t}^2 } } \> ,
\end{equation}
\begin{equation}
H= \sum_{j=1}^2 {\bf p}_j \cdot {\bf v}_j - L= \sum_{j=1}^2
{m_j \over \sqrt{ 1 -{\bf v}_j^2}} + \sigma r \int_0^1 ds {1\over
\sqrt{1- {\bf v}_{\rm t}^2}} \> .
\end{equation}
Here we have set ${\bf v}_j =\dot{\bf z}_j$ and
by ${\bf v}_{\rm t}(s) =s \, \dot{\bf z}_{1{\rm T}}
+ (1-s) \dot{\bf z}_{2{\rm T}}$ we have denoted  the velocity of the
flux tube segment specified by $s$ and $s +ds$.\par
Eqs. (15) cannot be inverted in closed form, but  it can be by an
expansion  in $1 /m^2$ or in $\sigma$. In the first case we
reobtain the semirelativistic Wilson-loop potential [1] (apart from
the spin-dependent terms), in the second one we find [4]
at the first order in $\sigma$
\begin{eqnarray}
H && = \sqrt{m_1^2+q^2} +\sqrt{m_2^2+q^2}+
{\sigma r\over 2}
{1\over \sqrt{m_1^2+q^2} +\sqrt{m_2^2+q^2}} \\
& & \times \left\{ \sqrt{m_2^2+q^2\over m_1^2+q^2} \sqrt{m_1^2+q_r^2}+
\sqrt{m_1^2+q^2\over m_2^2+q^2} \sqrt{m_2^2 +q_r^2} \right.
+\nonumber \\
& &+ \left. \left( {\sqrt{m_1^2+q^2} \sqrt{m_2^2+q^2} \over q_{\rm T}}
\right)  \left( {\rm arcsin} {q_{\rm T}\over \sqrt{m_1^2+q^2}}
+{\rm arcsin}{q_{\rm T}\over \sqrt{ m_2^2+q^2}} \right) \right\}
\> , \nonumber
\end{eqnarray}
where relevant simplifications have been obtained setting explicitly
${\bf p}_1= - {\bf p}_2 = {\bf q}$ (centre-of-mass frame) and where
we have defined ${\bf q}_r = ({\bf q}\cdot \hat{r}) \hat{r}$. \par
Notice that if we use (18) in the phase space path integral and try to go
back to Eq. (12) by integrating again over the momenta, we would
find a different $\Delta({\bf z})$  due to the occurrence in (18)
of a momentum-dependent interaction part. This means that
to be consistent  we have actually to introduce an appropriate
normalization factor in front of the expression of $W_{q \bar{q}}$
as would be given by (8) (see [7]).

\section{Three-quark flux tube Lagrangian}

The three-quark gauge invariant  propagator is
\begin{eqnarray}
& &G(x_1,x_2,x_3,y_1,y_2,y_3)
=\nonumber\\
& &\quad\quad\quad\quad
=\frac{1}{3!} \varepsilon_{a_1 a_2 a_3} \varepsilon_{b_1 b_2 b_3}
\nonumber\\
& &\langle 0| {\rm T} \,
U^{a_3 c_3}(x_M,x_3) U^{a_2 c_2}(x_M,x_2)U^{a_1 c_1}(x_M,x_1)
\psi_{3 c_3}(x_3)\psi_{2 c_2}(x_2) \psi_{1 c_1}(x_1)
\nonumber\\
& &\quad \> \overline{\psi}_{1 d_1}(y_1) \overline{\psi}_{2 d_2}(y_2)
\overline{\psi}_{3 d_3}(y_3)U^{d_1 b_1}(y_1,y_M) U^{d_2 b_2}(y_2,y_M)
U^{d_3 b_3}(y_3,y_M) |0 \rangle
\nonumber\\
& &\quad\quad\quad\quad
=\frac{1}{3!} \varepsilon_{a_1 a_2 a_3}\varepsilon_{b_1 b_2 b_3}
\nonumber\\
& &\bigg\langle\>
        \bigg(U(x_M,x_1)S_1^{\rm F}(x_1,y_1|A)U(y_1,y_M)\bigg)^{a_1 b_1}
\nonumber\\
& &\quad\bigg(U(x_M,x_2)S_2^{\rm F}(x_2,y_2|A)U(y_2,y_M)\bigg)^{a_2 b_2}
\nonumber \\
& &\quad\bigg(U(x_M,x_3)S_3^{\rm F}(x_3,y_3|A)U(y_3,y_M)\bigg)^{a_3 b_3}
\bigg\rangle \> ,
\end{eqnarray}
where we assume $x_1^0=x_2^0=x_3^0=x_M^0=t_{\rm f}$,
$y_1^0=y_2^0=y_3^0=y_M^0=t_{\rm i}$, ${\bf x}_M$ and $ {\bf y}_M$
are points chosen inside the triangles $({\bf x}_1, {\bf x}_2,
{\bf x}_3)$ and $({\bf y}_1, {\bf y}_2, {\bf y}_3)$ in such a way
that $\sum_{j=1}^3 \vert{\bf x}_j- {\bf x}_M\vert$
and $\sum_{j=1}^3 \vert {\bf y}_j -{\bf y}_M\vert $ are minima
and $a_i$, $b_i$ are colour indices. \par
Again, neglecting spin-independent terms, we can write
\begin{eqnarray}
& &K({\bf x}_1, {\bf x}_2, {\bf x}_3; {\bf y}_1, {\bf y}_2,
{\bf y}_3; t_{\rm f} - t_{\rm i}| \, {\bf x}_M , {\bf y}_M )  =
\nonumber\\
& & \quad\quad\quad
= \int_{{\bf y}_1}^{{\bf x}_1} {\cal D} {\bf z}_1 {\cal D}{\bf p}_1
\int_{{\bf y}_2}^{{\bf x}_2} {\cal D} {\bf z}_2 {\cal D}{\bf p}_2
\int_{{\bf y}_3}^{{\bf x}_3} {\cal D} {\bf z}_3 {\cal D}{\bf p}_3
\nonumber\\
& & \quad\quad\quad\quad
\exp \left\{ i \left[ \int_{t_{\rm i}}^{t_{\rm f}} dt \, \sum_{j=1}^{3}
\left( {\bf p}_j \cdot \dot{{\bf z}}_j - \sqrt{m_j^2 +{\bf p}_j^2}
\right) -i \ln W_{3q} \right] \right\}  \> ,
\end{eqnarray}
which corresponds to (7).
$W_{3q}$ is the ``Wilson loop integral'' for three quarks defined by
\begin{eqnarray}
W_{3q} &=& \frac{1}{3!} \left\langle \varepsilon_{a_1 a_2 a_3}
 \varepsilon_{b_1 b_2 b_3}  \left[ {\rm  P} \exp \left( ig
 \int_{\overline{\Gamma}_1} dx^{\mu_1} A_{\mu_1}(x) \right)
\right]^{a_1 b_1} \right. \\
& & \left. \left[ {\rm P} \exp \left( ig \int_{\overline{\Gamma}_2}
dx^{\mu_2} A_{\mu_2}(x) \right) \right]^{a_2 b_2}
  \left[  {\rm P} \exp \left( ig \int_{\overline{\Gamma}_3 } dx^{\mu_3}
A_{\mu_3}(x) \right) \right]^{a_3 b_3} \right\rangle \> ,
\nonumber
\end{eqnarray}
$\bar{\Gamma}_j$ being made by the world line $\Gamma_j$ of the quark $j$
plus the straight lines connecting $y_M$ to $y_j$ and $x_j$ to $x_M$
(see Fig. 2). As in (8) we shall set
\begin{equation}
i  \ln W_{3q}= \frac{2}{3} g^2 \sum _{i<j}
 \int _{\Gamma _i} dx^{\mu}
_i \int _{\Gamma _j} dx^{\nu}_j \,i D_{\mu \nu} (x_i - x_j)
+ \sigma S_{\min} +  C P  \>,
\label{aar}
\end{equation}
where the first term is again the lowest-order  perturbative contribution;
${ P}$ is  the total length of $\bar{\Gamma}_1$, $\bar{\Gamma}_2$ and
$\bar{\Gamma}_3$, $S_{\rm min}$ is the area of the minimal three-sheet
surface enclosed by $\bar{\Gamma}_1$, $\bar{\Gamma}_2$, $\bar{\Gamma}_3$
and a mid-point world-line $\Gamma_M$ ($z_M= z_M(t)$) joining
$y_M$ and $x_M$ ; the minimum is performed for fixed quark world
line, but varying $\Gamma_M$.\par
Again adopting instantaneous and straight line approximations we obtain
\begin{eqnarray}
i  \ln W_{3 q} = && \int_{t_{\rm i}}^{t_{\rm f}} dt \, \left\{
 \sum_{j<l} {-2 \over 3} { {\alpha}_s  \over r_{jl}}
\left[ 1 - \frac{1}{2}
( \delta^{hk} + \hat{r}^h_{jl} \hat{r}^k_{jl} )
\dot{z}_j^h \dot{z}_l^k \right] + \right.
\nonumber\\
{} &+& \sum_{j=1}^3 \sigma \, r_j \int_0^1 ds_j
\left[1 - \left( s_j \dot{\bf z}_{j{\rm T}_j} + (1-s_j) \dot{{\bf
z}}_{M{\rm T}_j} \right)^2 \right]^{1/2} +
\nonumber\\
{} &+& \left. C \sum_{j=1}^3 \sqrt{1-\dot{\bf z}_j^2} \right\}
 \>.
\end{eqnarray}
Here ${\bf r}_{ij} = {\bf z}_i -{\bf z}_j$,
${\bf r}_j= {\bf z}_j - {\bf z}_M$ and ${\bf z}_M$
has to be chosen  in such a way that $\sum_{j=1}^3 {r}_j$ is minimum
at any given time, $\dot{z}_{j{\rm T}_j}^h = (\delta^{hk} - \hat{r}_j^h
\hat{r}_j^k) \dot{z}_j^k $.
Of course, ${\bf z}_M(t)$ has to be coplanar with ${\bf z}_1(t)$,
${\bf z}_2(t)$ and ${\bf z}_3(t)$ and then two different types of
configurations are possible:
\par\noindent
I)\,\, if in the triangle (${\bf z}_1(t)$, ${\bf z}_2(t)$, ${\bf z}_3(t)$)
 no angle exceeds  $120^0$, then ${\bf z}_M(t)$ has to be such that
 ${\bf r}_1(t)$, ${\bf r}_2(t)$, ${\bf r}_3(t)$ make angles of $120^0$
with each other;\par\noindent
II) if the angle in ${\bf z}_l(t)$ reaches $120^0$, then
 ${\bf z}_M= {\bf z}_l$.\par
 Notice that correspondingly for $\dot{\bf z}_M$ we have
\begin{equation}
\dot{\bf z}_{M} = \left\{
\begin{array} {ll}
{\cal R}^{-1} \sum_{j=1}^3 \left( \dot{\bf z}_{j {\rm T}_j}/r_j   \right)
\qquad\quad \mbox{I type configuration}& \\
\dot{\bf z}_l  \qquad \qquad
\qquad \qquad \qquad \> \mbox{ II type configuration}
\end{array} \right.
\end{equation}
${\cal R}$ being the matrix with elements
${\cal R}^{hk}= \sum_{j=1}^3 (\delta^{hk} -\hat{r}^h_j \hat{r}^k_j)/r_j$.\par
Replacing as before (23) in (20) and integrating on the momenta in the Gaussian
approximation we can write
\begin{eqnarray}
K({\bf x}_1, {\bf x}_2, {\bf x}_3; {\bf y}_1, {\bf y}_2,
{\bf y}_3; t_{\rm f} - t_{\rm i}) &=&
\int_{{\bf y}_1}^{{\bf x}_1} {\cal D} [ {\bf z}_1 ] \Delta [ {\bf z}_1 ]
\int_{{\bf y}_2}^{{\bf x}_2} {\cal D} [ {\bf z}_2 ] \Delta [ {\bf z}_2 ]
\int_{{\bf y}_3}^{{\bf x}_3} {\cal D} [ {\bf z}_3 ] \Delta [ {\bf z}_3 ]
\nonumber\\
& &\exp \left\{i \int_{t_{\rm i} }^{t_{\rm f}}
dt \, L( {\bf z}_1, {\bf z}_2, {\bf z}_3,
\dot{\bf z}_1, \dot{\bf z}_2, \dot{\bf z}_3 ) \right\} \>,
\label{adq}
\end{eqnarray}
and obtain the three-quark flux tube Lagrangian
\begin{eqnarray}
L = &-& \sum_{j=1}^3 m_j \sqrt{1- \dot {\bf z}_j^2 } +
\sum_{j<l} {2 \over 3} { {\alpha}_s  \over r_{jl}}
\left[ 1 - \frac{1}{2}
\left( \delta^{hk} + \hat{r}^h_{jl} \hat{r}^k_{jl} \right)
\dot{z}_j^h \dot{z}_l^k \right] -
\nonumber\\
{} &-& \sum_{j=1}^3 \sigma \, r_j \int_0^1 ds_j
\left[1 - \left( s_j \dot{\bf z}_{j{\rm T}_j} + (1-s_j) \dot{{\bf
z}}_{M{\rm T}_j} \right)^2\right]^{1/2} -
\nonumber\\
{} &-& {C} \sum_{j=1}^3 \sqrt{1-\dot{\bf z}_j^2} \>.
\label{adr}
\end{eqnarray}
Notice that again the perimeter term can be absorbed in a redefinition
of the  quark masses.

\section{\bf Three-quark relativistic flux tube Hamiltonian }

For simplicity, in this section we shall omit the perturbative part
of the Lagrangian, keeping in mind that this can be simply added to the
pure  flux tube result.
To simplify the notations let us set ${\bf v}_i=\dot{\bf z}_i$;
${\bf v}_M= \dot{\bf z}_M$, ${\bf v}_i^t(s_i) = s_i {\bf v}_{iT_{i}}
+ (1-s_i) {\bf v}_{MT_i}$. Let us also introduce the matrices
${\cal T}_i$ with elements ${\cal T}_i^{hk}=\delta^{hk} -\hat{r}_i^h
\hat{r}_i^k $; then we have obviously
\begin{equation}
{\cal R} = \sum_{j=1}^3 {1\over r_j} {\cal T}_j \>,
\end{equation}
and
\begin{eqnarray}
& &{\bf v}_{i T_i} = {\cal T}_i {\bf v}_i \>, \quad
{\bf v}_M = {\cal R}^{-1} \sum_{j=1}^3 {1\over r_j} {\cal T}_j {\bf v}_j
\>,
\nonumber \\
& &{\bf v}_i^t(s_i) = {\cal T}_i (s_i {\bf v}_i + (1-s_i) {\bf v}_M) =
s_i{\cal T}_i {\bf v}_i + (1-s_i) \sum_{j=1}^3 {1 \over r_j} \,
{\cal T}_i {\cal R}^{-1} {\cal T}_j \, {\bf v}_j \>.
\end{eqnarray}
So we can write the flux tube Lagrangian as
\begin{equation}
L = -\sum_{i=1}^3  m_i \sqrt{1-{\bf v}_i^2}
- \sigma \sum_{i=1}^3 r_i\int_0^1 ds_i \sqrt{1-{\bf v}^{t\,2}_{i}} \>
\end{equation}
and, we can obtain the
canonical conjugate momenta
\begin{equation}
{\bf p}_i = { m_i {\bf v}_i \over \sqrt{1-{\bf v}_i^2} }
+ \sigma \sum_{j=1}^3 r_j\int_0^1 ds_j
{\left( s_j \delta_{ij} + (1-s_j){\cal R}^{-1}
{\cal T}_i /r_i \right) {\bf v}^t_{j}
\over \sqrt{1-{\bf v}^{t\,2}_{j}} } \>,
\end{equation}
while the total linear momentum and the Hamiltonian turn out,
respectively, to be:
\begin{equation}
{\bf P} = \sum_{i=1}^3 {\bf p}_i
= \sum_{i=1}^3 { m_i {\bf v}_i \over \sqrt{1-{\bf v}_i^2} }
+ \sigma \sum_{i=1}^3 r_i\int_0^1 ds_i
{ {\bf v}^t_{i} \over \sqrt{1-{\bf v}^{t\,2}_{i}} }  \> ,
\end{equation}
\begin{equation}
H  =  \sum_{i=1}^3 {\bf p}_i \cdot {\bf v}_i - L
= \sum_{i=1}^3 { m_i \over \sqrt{1-{\bf v}_i^2} }
+ \sigma \sum_{i=1}^3 r_i
\int_0^1 ds_i { 1 \over \sqrt{1- {\bf v}^{t\,2}_{i}} } \>.
\end{equation}
Again Eq. (30) can be solved with respect to ${\bf v}_j$ ($j=1,2,3$)
only as an expansion in $1/m_j^2$ or in the string tension $\sigma$.
Consequently even $H$ turns out to be expressed in such a form.\par
Using an expansion in $1/m_j^2$ we obtain
\begin{eqnarray}
H  &=& \sum_{i=1}^3 m_i
+ {1\over 2} \sum_{i=1}^3 {{\bf p}^2_i \over m_i}
- {1\over 8} \sum_{i=1}^3 {{\bf p}^4_i \over m_i^3} -
\nonumber\\
&-& {\sigma \over 6} \sum_{i=1}^3 r_i \left.
\left(  {{\bf p}_{iT_i}^2
\over m_i^2} + {\bf v}_{MT_i}^2
+ {{\bf p}_{iT_i} \cdot {\bf v}_{MT_i}\over m_i} \right)
\right|_{{\bf v}_i = {\bf p}_i/m_i} + O\left( {\sigma^2 \over m^3} \right)
\> ,
\end{eqnarray}
with
\begin{equation}
\left. {\bf v}_{MT_i} \right|_{{\bf v}_i = {\bf p}_i/m_i}
= {\cal R}^{-1} \sum_j {1\over r_j}  {\cal T}_j
{{\bf p}_{j{\rm T}_i}\over m_j} \>.
\end{equation}
Eq. (33) coincides with the static plus the velocity-dependent part of the
three-quark Wilson loop Hamiltonian [1].\par
Considering an expansion in $\sigma$ in (30) we obtain the
explicit relativistic Hamiltonian
\begin{equation}
H = \sum_{i=1}^3 \sqrt{p_i^2+m_i^2}
+ \sigma \sum_{i=1}^3 r_i \left. \int_0^1 ds_i
\, \sqrt{1- {\bf v}^{t\,2}_{i}} \,
\right|_{{\bf v}_i = {\bf p}_i/\sqrt{p_i^2 +m_i^2}}
+ O\left( \sigma^2 \right) \>,
\end{equation}
obviously now
\begin{equation}
{\bf v}_i^t\vert_{{\bf v}_i = {{\bf p}_i\over \sqrt{m_i^2+{\bf
p}_i^2}}} = s_i{{\bf p}_{i {\rm T}_i }\over \sqrt{m_i^2 + {\bf p}_i^2}}
+ (1-s_i) {\cal R}^{-1} \sum_{j=1}^3
{1\over r_j}{\cal T}_j {{\bf p}_{j {\rm T}_i} \over \sqrt{m_j^2+{\bf p}_j^2}}
\> .
\end{equation}
Notice that like Eq. (9), even Eq. (23) and so even Eqs. (26), (29), (32) are
supposed to hold in the centre-of-mass frame ${\bf P}={\bf p}_1
+{\bf p}_2 + {\bf p}_3 =0 $ and ${\bf p}_1, {\bf p}_2 ,{\bf p}_3 $
 are not independent. A possible choice of independent variables  is
given  by the internal Jacobi momenta, in terms of which
we can write
\begin{equation}
{\bf p}_1 = {\bf q} \>, \quad \quad
{\bf p}_2 = - { m_2 \over m_2 +m_3 } {\bf q}+ {\bf k} \>, \quad \quad
{\bf p}_3 = -{m_3 \over m_2 + m_3 } {\bf q} - {\bf k} \>.
\end{equation}
Eq. (37) should be used in (36), (35). The final result
is very complicated. For low angular momenta, however, $ v_i^{t\,2}$
should be negligible and Eq. (35) reduces to the three-quark starlike
string model
\begin{equation}
H= \sum_{i=1}^3 \sqrt{m_i^2 + {\bf p}_i^2 } +\sigma \sum_{i=1}^3 r_i
\> ,
\end{equation}
already considered in the literature [5,6].\par
Performing explicitly the  $s_i$ integration in (35) we obtain
\begin{eqnarray}
H &=&  \sum_{i=1}^3 \sqrt{p_i^2+m_i^2}+
\sigma  \sum_{i=1}^3 r_i \Biggr\{ {1\over 2}
\sqrt{1-{\bf v}_{i{\rm T}_i}^2} +
\nonumber \\
&+& {{\bf v}_{M{\rm T}_i} \cdot
({\bf v}_{i{\rm T}_i} - {\bf v}_{M{\rm T}_i}) \over 2 ({\bf v}_{i{\rm T}_i}-
{\bf v}_{M{\rm T}_i})^2 } \left[ \sqrt{1-{\bf v}_{i{\rm T}_i}^2}
- \sqrt{1-{\bf v}_{M{\rm T}_i}^2} \right] \nonumber \\
&+& {1\over 2 \vert {\bf v}_{i{\rm T}_i} - {\bf v}_{M{\rm T}_i}\vert }
 \left[ (1- {\bf v}_{M{\rm T}_i}^2) + {
({\bf v}_{i{\rm T}_i} \cdot
({\bf v}_{i{\rm T}_i} - {\bf v}_{M{\rm T}_i}))^2 \over
({\bf v}_{i{\rm T}_i} - {\bf v}_{M{\rm T}_i})^2 } \right]
\nonumber \\
&\times& \left( {\rm arcsin}{ {\bf v}_{i{\rm T}_i} \cdot
 ({\bf v}_{i{\rm T}_i} - {\bf v}_{M{\rm T}_i}) \over
 \sqrt{ (1- {\bf v}_{M{\rm T}_i}^2)
({\bf v}_{i{\rm T}_i}- {\bf v}_{M{\rm T}_i})^2 +
({\bf v}_{i{\rm T}_i} \cdot
 ({\bf v}_{i{\rm T}_i} - {\bf v}_{M{\rm T}_i}))^2}} \right.
\nonumber \\
& & + \left. {\rm arcsin}
{ {\bf v}_{M{\rm T}_i} \cdot
 ({\bf v}_{i{\rm T}_i} - {\bf v}_{M{\rm T}_i}) \over
 \sqrt{ (1- {\bf v}_{M{\rm T}_i}^2)
({\bf v}_{i{\rm T}_i}- {\bf v}_{M{\rm T}_i})^2 +
({\bf v}_{i{\rm T}_i} \cdot
 ({\bf v}_{i{\rm T}_i} - {\bf v}_{M{\rm T}_i}))^2}} \right) \Biggr\}
\end{eqnarray}
which, however, does not essentially simplify after the substitution (37),
contrary to  what happens in the quark-antiquark case (18).

\section{\bf Conclusion}

In conclusion we have shown that even for the three-quark system
a flux-tube like relativistic Lagrangian can be obtained in the
Wilson-loop framework, if one neglects spin. The method
does not seem appropriate for a direct introduction of spin since it uses
a Foldy--Wouthuysen type transformation at an early stage and so
the spin dependence in the starting path-integral occurs as an expansion
in $1/m^2$. \par
As a matter of fact, in the quark-antiquark case the relativistic flux-tube
model is strictly related to the instantaneous approximation of the
corresponding Bethe--Salpeter equation, and spin can be taken into account
through such an equation [8]. In principle the generalization of such an
equation should be the convenient formalism even for the three-quark
bound state. \par
Notice finally that to give the quantum Hamiltonian operator corresponding
to Eqs. (35) and (39), attention has to be paid to questions of
ordering. If the discrete form of the quantity $W_{3q}$ is written
as [1]
\begin{eqnarray}
W_{3q}&=& {1\over 3!} \Biggr\langle
\varepsilon_{a_1 a_2 a_3}\varepsilon_{b_1 b_2 b_3}
\\
& & \prod_{j=1}^3 \left[ {\rm P} \, \exp \left( i \, g \sum_{\Gamma_j}
\left( z^{\mu}_{j \, n} - z^{\mu}_{j \, n-1} \right)
A_{\mu}\left( {z_{j\,n} + z_{j\,n-1} \over 2} \right) \right)
\right]^{a_j b_j}  \Biggr\rangle \>,
\nonumber
\end{eqnarray}
the correct ordering would be the Weyl's ordering.

\newpage
\begin{figure}
\vspace{15.5truecm}
\caption{Quark-antiquark Wilson loop}
\end{figure}
\vfill\eject

\newpage
\begin{figure}
\vspace{15.5truecm}
\caption{Three-quark Wilson loop}
\end{figure}
\end{document}